\documentclass[pre,twocolumn]{revtex4}
\usepackage{epsfig,graphics}
\usepackage{float}
\newcommand{\br}{{\bf r}}

\newcommand{\bR}{{\bf R}}

\newcommand{\be}{\begin{equation}}
\newcommand{\ee}{\end{equation}}

\newcommand{\Sex}{s_{\rm ex}}
\newcommand{\sex}{s_{\rm ex}}

\begin{document}

\title{Estimating the density-scaling exponent of a monatomic liquid from its pair potential}
\author{Lasse B{\o}hling, Nicholas P. Bailey, Thomas B. Schr{\o}der, and Jeppe C. Dyre}
\email{dyre@ruc.dk}
\affiliation{DNRF Center ``Glass and Time'', IMFUFA, Dept. of Sciences, Roskilde University, P. O. Box 260, DK-4000 Roskilde, Denmark}
\date{\today}

\begin{abstract}
This paper investigates two conjectures for calculating the density dependence of the density-scaling exponent $\gamma$ of a single-component, pair-potential liquid with strong virial potential-energy correlations. The first conjecture gives an analytical expression for $\gamma$ directly in terms of the pair potential. The second conjecture is a refined version of this involving the most likely nearest-neighbor distance determined from the pair-correlation function. The conjectures are tested by simulations of three systems, one of which is the standard Lennard-Jones liquid. While both expressions give qualitatively correct results, the second expression is more accurate.
\end{abstract}

\maketitle

\section{Introduction}\label{intro}

Temperature is the standard parameter varied in experiments investigating a liquid's structure, dynamics, and thermodynamics. The thermodynamic phase diagram is not one- but two-dimensional, however, so liquid properties can only be mapped out completely by probing also high-pressure states. In the study of glass-forming liquids, in particular, high-pressure experiments have led to important new insights. Thus in the last decade a regularity termed ``density scaling'' has been convincingly established for a large class of glass-forming liquids \cite{tol01,alb04,dre04b,rol05,rol10,she12,grz13,xu13}, a scaling that also applies for less-viscous ``ordinary'' liquids if proper reduced units are used \cite{fra11}. If $T$ is the temperature and $\rho$ the density, a viscous liquid obeys density scaling if its relaxation time -- or, equivalently, viscosity -- is a function of $\rho^\gamma/T$ for some exponent $\gamma$. An important insight from the discovery of density scaling is that density, not pressure, is the relevant thermodynamic variable for understanding the dynamics of liquids. The most widely investigated systems in experiments are organic liquids and polymers. It has been found that van der Waals bonded systems obey density scaling to a good approximation, whereas hydrogen-bonded liquids like glycerol often disobey density scaling \cite{rol05,rol08,fra10}. 

The isomorph theory provides a theoretical framework for understanding the origin of density scaling for a large class of systems \cite{IV,dyr13a}. According to this theory, a liquid obeys density scaling whenever it has strong correlations between its virial and potential-energy thermal equilibrium fluctuations at constant volume. We originally called such liquids ``strongly correlating''. Many people inferred a connection to strongly correlated quantum liquids, however, so we now refer to the relevant class as ``Roskilde-simple liquids'', which reflects the fact that these liquids are in many respects simpler than other liquids \cite{ing12} (the term ``simple liquid'' is traditionally used for all monatomic pair-potential systems, but some of them do not have strong correlations whereas many molecular systems do). It appears that most or all van der Waals and metallic liquids are Roskilde simple, whereas covalently- and hydrogen-bonded liquids, due to their directional bonding, are not. Likewise, systems with strong ionic or dipolar interactions are generally not strongly correlating, but systems with weaker such interactions may well be. Much more work is needed to get the full overview of the class of Roskilde-simple systems. 

A Roskilde-simple liquid has isomorphs in its thermodynamic phase diagram. These are curves along which a number of structural, dynamic, and thermodynamic properties are invariant in reduced units \cite{IV}. In particular, the excess entropy $\Sex$ (the entropy minus that of an ideal gas at same density and temperature) is an isomorph invariant. Since the excess entropy is the entropy of the configurational degrees of freedom, isomorphs are configurational adiabats. The opposite does not apply, however, because all systems have configurational adiabats.

According to the isomorph theory the density-scaling exponent $\gamma$ generally varies with the thermodynamic state point, but only as a function of the density: $\gamma=\gamma(\rho)$. The simulations presented below confirm that density is the dominating factor. In experiments density often does not vary much (5-10\%) and the assumption of a constant $\gamma$ usually works well \cite{rol05}. Recently it was shown, however, that for larger density variations $\gamma$ is not constant \cite{boh12}. Isomorph scaling applies also in this more general case \cite{boh12}.

The present paper extends and tests recent results of ours \cite{bai13} on approximations for calculating the density-scaling exponent  of single-component systems with pairwise additive interactions. In contrast to the case of molecular liquids for which there still is no theory for $\gamma$, we show that one can arrive at a reasonably good understanding of the scaling properties of monatomic pair-force liquids. Section \ref{theo} gives the necessary theoretical background and arrives at two approximate expressions for the density-scaling exponent, Sec. \ref{comp} presents the three systems studied numerically, and Sec. \ref{compar} compares simulation results to the predictions of the two approximations. Finally, Sec. \ref{disc} shows that the approximation are equivalent to postulating isomorph invariance of the effective Einstein frequency of the pair interaction at a certain distance.

\section{Theoretical background}\label{theo}

This section gives the background and motivation for the simulations presented in the next section. Much of the material in Secs. \ref{bas} and \ref{ds_sec} is based on previous papers (Refs. \onlinecite{II,ing12,ing12a}) and may be skipped by readers thoroughly familiar with these works. Section \ref{eipl_sec} arrives at the two conjectures tested in the simulations.

\subsection{Basics}\label{bas}

We consider a classical-mechanical system of $N$ particles of mass $m$ in volume $V$ with density $\rho\equiv\ {N}/{V}$. If the particle positions are denoted by $\br_1,...,\br_N$, the collective $3N$-dimensional position vector is defined by $\bR\equiv (\br_1,...,\br_N)$. ``Reduced units'' refer to the unit system in which the length unit is $\rho^{-1/3}$, the energy unit is $k_BT$ where $k_B$ is Boltzmann's constant, and the time unit is $\rho^{-1/3}\sqrt{m/k_BT}$. 

By uniform scaling of all coordinates a given microconfiguration of a thermodynamic state point corresponds to a microconfiguration at another density. By definition \cite{IV}, two state points $(\rho_1,T_1)$ and $(\rho_2,T_2)$ are isomorphic if a constant $C_{12}$ exists such that the following applies: whenever two physically relevant microconfigurations of the state points, $\bR_1\in(\rho_1,T_1)$ and $\bR_2\in(\rho_2,T_2)$, have the same reduced coordinates, i.e., $\rho_1^{1/3}\bR_1=\rho_2^{1/3}\bR_2$, one has

\be\label{iso_def1}
\exp\left(-\frac{U(\bR_1)}{k_BT_1}\right)
\,=\,C_{12}\exp\left(-\frac{U(\bR_2)}{k_BT_2}\right)\,.
\ee
This defines a mathematical equivalence relation in the thermodynamic phase diagram, the equivalence classes of which are the system's ``isomorphs''. It is straightforward to show that for an Euler-homogeneous potential-energy function of order $-n$, two state points are isomorphic whenever $\rho_1^{n/3}/T_1=\rho_2^{n/3}/T_2$. In this case $C_{12}=1$, but for realistic systems $C_{12}\neq 1$ (in which case the reduced-unit free energy varies along an isomorph).

The identity Eq. (\ref{iso_def1}) implies that several quantities are isomorph invariant when given in reduced units \cite{IV}. Examples are the excess entropy, the isochoric specific heat, the instantaneous shear modulus, the diffusion constant, the viscosity, etc. In fact, the entire reduced-unit microscopic dynamics is predicted to be invariant along an isomorph, and so are all structural measures, including higher-order spatial correlation functions. Of course, since isomorphs are approximate, isomorph invariance is not exact. 

Recall that the virial $W(\bR)\equiv (-1/3) \bR \cdot {\bf \nabla} U(\bR)$ gives the contribution to the pressure $p$ from the interactions \cite{tildesley,han13}, which means that the average virial $\langle W\rangle$ modifies the ideal-gas equation of state into $pV=Nk_BT+\langle W\rangle$. The Pearson correlation coefficient $R$ of the equilibrium, constant-volume $WU$ fluctuations is defined by (where the brackets denote constant-volume canonical averages)

\be\label{R}
R \,=\,
\frac{\langle\Delta W\Delta U\rangle}
{\sqrt{\langle(\Delta W)^2\rangle\langle(\Delta U)^2\rangle}}\,.
\ee
The criterion $R>0.9$ provides a pragmatic delimitation of the class of Roskilde-simple liquids \cite{ped08}.

Few if any systems with attractions have isomorphs in their entire phase diagram. Computer simulations have shown \cite{IV,ped10,ing12,ing12a,ing12b,vel12,vel13} that a typical Roskilde-simple liquid has good isomorphs throughout its condensed liquid phase, in fact, including the entire crystalline phase \cite{cryst}. When the critical point and the gas phase are approached, however, the isomorph theory breaks down. High-pressure, high-temperature supercritical state points have good isomorphs when these are not too far away from the solid-liquid coexistence curve. Incidentally, this curve is an isomorph, a fact that explains the many invariants along the melting curve identified throughout the years (see, e.g., Refs. \onlinecite{IV,V,dyr13} and their references).

Several liquids have been found in simulations to be Roskilde-simple, for instance \cite{I,ped08,ing12,ing12b,vel12,vel13}: The Lennard-Jones (LJ) system \cite{lj24} and its generalizations to mixtures and to other exponents than 6 and 12, simple molecular liquids like the asymmetric dumbbell or the Lewis-Wahnstr{\"o}m OTP model \cite{otp2}, the Buckingham liquid with an exponential repulsive term \cite{buc38}, the ``Repulsive'' LJ system (with plus instead of minus between the two terms) \cite{ing12a}. Recently it was shown that even the 10-bead rigid-bond, flexible LJ chain has good isomorphs \cite{vel13}, providing a highly nontrivial example of a Roskilde-simple liquid. As mentioned, the theory works well for the crystalline phase; thus a (classical) LJ crystal has $R>0.99$ \cite{II,cryst}. In all cases the theory was checked by tracing out isomorphs in the thermodynamic phase diagram and testing for the predicted invariants. The different methods that can be used for generating isomorphs in simulations have been detailed elsewhere \cite{IV,ing12a,boh13}. 

Roskilde-simple liquids have simple thermodynamics. If $\sex$ is the excess entropy per particle, temperature factorizes as follows \cite{ing12a} $k_BT=f(\sex)h(\rho)$. Since excess entropy is an isomorph invariant, the isomorphs are consequently given by

\be\label{isomch}
\frac{h(\rho)}{k_BT}={\rm Const.}
\ee

\subsection{Defining the density-scaling exponent}\label{ds_sec}

The density-scaling exponent $\gamma$ is defined here by \cite{IV} 

\be\label{gammadef}
  \gamma \equiv  \left(\frac{\partial \ln T}{\partial \ln \rho} \right)_{\Sex}\,.
\ee
In experiment one would define $\gamma$ by keeping not the excess entropy, but the relaxation time constant, defining a so-called isochrone in the thermodynamic phase diagram. In practice there is little difference between these definitions, because according to the isomorph theory both entropy and reduced relaxation time are constant along an isomorph \cite{IV}. For the systems for which density scaling has been most thoroughly studied in experiment -- supercooled liquids and polymers -- the difference between real and reduced relaxation time is insignificant because the dramatic density and temperature dependence of the relaxation time totally dominates over the factor multiplied by in order to switch to reduced units. Recent works including also data for less-viscous liquids show, however, the importance of working with reduced units to get proper density-scaling exponents \cite{fra11,fra11a}.

Whenever the right-hand side of Eq. (\ref{gammadef}) is constant, the isomorphs are given by the well-known density-scaling expression $\rho^\gamma/T=$Const. \cite{rol05}. As mentioned, the density-scaling exponent is usually identified in experiments by tracing out the isochrones, i.e., from the dielectric loss-peak frequency's variation with temperature and density \cite{rol05}. In computer simulations $\gamma$ is identified from the $NVT$ fluctuations using the identity \cite{IV}

\be \label{gamma_eq}
\gamma= \frac{\langle \Delta U \Delta W \rangle}{\langle(\Delta U)^2\rangle}\,.
\ee

If $v(r)=\sum_n \varepsilon_n (r/\sigma)^{-n}$, the function $h(\rho)$ inherits this analytical structure in the sense that

\be\label{heq}
h(\rho)=\sum_n \alpha_n \varepsilon_n(\rho \sigma^3)^{n/3}
\ee
is a sum over the same $n$ as appearing in $v(r)$ (where $\alpha _n$ are constants) \cite{ing12a,boh12}. In combination with Eq. (\ref{isomch}) this provides a convenient recipe for tracing out isomorphs in the phase diagram. Moreover, this provides an expression for $\gamma(\rho)$ because Eqs. (\ref{isomch}) and (\ref{gammadef}) imply

\be\label{gheq}
\gamma=\frac{d\ln h}{d\ln\rho}\,.
\ee
The analyticity property of $h(\rho)$ do not allow for a unique determination of $\gamma(\rho)$ from $v(r)$, however, because (except for inverse-power law (IPL) systems) the function $h(\rho)$ involves one or more parameters determined from simulations \cite{ing12a,boh12}. It is the purpose of the present paper to investigate to which extent one can estimate $\gamma(\rho)$ from $v(r)$.

\subsection{The eIPL approximation and two conjectures for estimating the density-scaling exponent}\label{eipl_sec}

For an inverse-power-law (IPL) pair potential, $v(r)=\varepsilon (r/\sigma)^{-n}$, the density-scaling exponent is constant throughout the phase diagram and given by

\be\label{gamipl}
\gamma=\frac{n}{3}\,.
\ee
This follows from Eqs. (\ref{heq}) and (\ref{gheq}) since $h(\rho)\propto\rho^{n/3}$. The question is how to generalize this result to realistic potentials. One way ahead starts from the fact that for an IPL pair potential $\propto r^{-n}$ the ratio of the $(p+1)$th and $p$th derivatives obeys $v^{(p+1)}(r)/v^{(p)}(r)=-(n+p)/r$. This reasoning led us in 2008 to define for any pair potential an effective, distance-dependent approximate IPL exponent $n^{(p)}(r)$ \cite{II} by

\be\label{np}
n^{(p)}(r)\equiv -p-r\frac{v^{(p+1)}(r)}{v^{(p)}(r)}\,,
\ee
which for the IPL case reduces to $n^{(p)}(r)=n$ for all $p$ and $r$. If one wishes to use Eq. (\ref{np}) for determining the density-scaling exponent via Eq. (\ref{gamipl}) for general pair potentials, the following questions arise: 1) Which value of $p$ to be used? 2) At which distance should $n^{(p)}(r)$ be evaluated? 3) How to relate this distance to the density? 

To address these questions we recall the ``extended IPL potential'' (eIPL) defined \cite{II} by $v_{\rm eIPL}(r)= a (r/\sigma)^{-n} + b + c (r/\sigma)$. As shown in Ref. \cite{II}, this potential gives an excellent fit to the Lennard-Jones (LJ) pair potential over the entire first coordination shell if one chooses an exponent $n\cong 18$. This is far from the value $n=12$ one would naively expect from the repulsive $r^{-12}$-term of the LJ potential. The reason is the often overlooked fact that due to the attractive term of the LJ pair potential, the repulsive {\it part} of the potential (i.e., below its minimum) is considerably steeper than predicted from the repulsive $r^{-12}$ {\it term} alone \cite{sti75,kan85,ben03,ped08,cos08,II}. At very high densities, or course, the physics is given by the $r^{-12}$-term.

In Ref. \cite{II} it was argued that the term $c (r/\sigma)$ contributes little to the fluctuations of virial and potential energy in the $NVT$ ensemble. The reason is that a given particle is surrounded by many others; if the particle is moved, some nearest-neighbor distances increase and others decrease, and the sum of all nearest-neighbor distances remains almost constant. Consequently, as regards the potential-energy fluctuations, the LJ system behaves largely as an IPL system with an exponent close to $18$; the same applies for the virial fluctuations \cite{II}. The eIPL approximation, however, misses the finer details of how and why the density-scaling exponent varies throughout the thermodynamic phase diagram; we here refer to it mainly as a source of inspiration.

Substituting $v_{\rm eIPL}(r)$ into Eq. (\ref{np}) for $p=2$ yields $n^{(2)}(r)=n$ for all $r$. With this in mind, the eIPL potential suggests using $n^{(2)}(r)$ for generally estimating the effective IPL exponent of a general pair potential of a Roskilde-simple liquid \cite{II}. 

At which distance should $n^{(2)}(r)$ be evaluated? One expects that the relevant distance is close to the typical nearest-neighbor distance. All distances scale with density as $\propto\rho^{-1/3}$, so we estimate the density-scaling exponent from (compare Eq. (\ref{gamipl}))

\be\label{gest1}
\gamma(\rho) = \left.\frac{ n^{(2)}(r)}{3}\right|_{r=\Lambda \rho^{-1/3}}\,.
\ee
It is straightforward to show that this expression implies the above-mentioned analyticity property of $h(\rho)$ when $v(r)=\sum_n \varepsilon_n (r/\sigma)^{-n}$. The value $\Lambda=2^{1/6}$ corresponds to the nearest-neighbor distance of an FCC crystal of density $\rho$, as pointed out by Lennard-Jones and Devonshire long ago \cite{lj37}. Equation (\ref{gest1}) is the first of the two expressions tested below by simulations.

There is no reason to believe, however, that the correct distance to use is the same for all thermodynamic state points \cite{bai13}. A more realistic choice is the distance corresponding to the most likely nearest-neighbor distance, i.e., the distance at which $r^2g(r)$ obtains its maximum, where $g(r)$ is the radial distribution function. Since structure is invariant along an isomorph, this implies the more general expression with $r=\Lambda(\Sex)\rho^{-1/3}$, i.e.,

\be\label{gest3}
\gamma(\rho,\Sex) = \left.\frac{ n^{(2)}(r)}{3}\right|_{r=\Lambda(\Sex)\rho^{-1/3}\equiv\rho_*^{-1/3}}\,.
\ee
In this case $\gamma$ is not a function exclusively of density as predicted by the isomorph theory \cite{IV}. Note that the dimensionless number $\Lambda(\Sex)$ is the reduced value of the $r$ giving the maximum of $r^2g(r)$.

\section{The systems studied by simulation}\label{comp}


In order to investigate how useful the above approximations are for estimating the density-scaling exponent for monatomic pair-potential liquids, we studied numerically three systems defined by the following pair potentials:

\begin{eqnarray}
  \label{eq:pots1}
  v_1(r) &=& 4\varepsilon \left \lbrace \left( \frac{r}{\sigma} \right )^{-12} - \left( \frac{r}{\sigma} \right )^{-6} \right \rbrace\\
  \label{eq:pots2}
  v_2(r) &=& \frac{\varepsilon}{2} \left \lbrace \left( \frac{r}{\sigma} \right )^{-12} + \left( \frac{r}{\sigma} \right )^{-6} \right \rbrace\\
  \label{eq:pots3}
  v_3(r) &=& \varepsilon \left \lbrace \left( \frac{r}{\sigma} \right )^{-18} - \left( \frac{r}{\sigma} \right )^{-12} + \left( \frac{r}{\sigma} \right )^{-6} \right \rbrace\,.
\end{eqnarray}
The negative term in $v_3(r)$ notwithstanding, the standard LJ potential $v_1(r)$ is the only one with attractive pair forces. These three systems were studied because, on the one hand, they are simple in their definition, while on the other hand they have qualitatively different behavior of the density-scaling exponent's variation with density (compare Figs. 4 and 5 below).

\begin{figure}[H]  
  \centering
  \includegraphics[trim = 0mm 0mm 0mm 0mm, clip,width= 0.3 \textwidth]{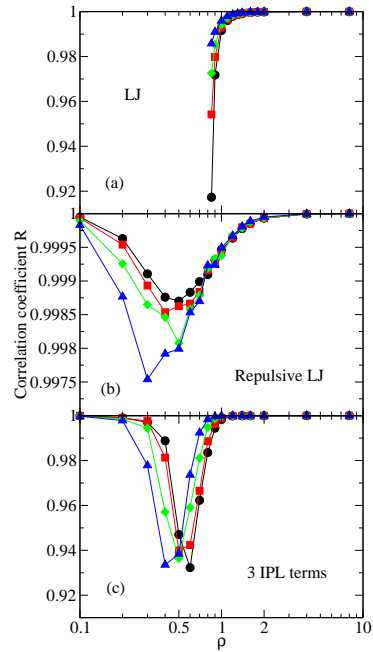}
  \caption{The virial potential-energy correlation coefficient $R$ of Eq. (\ref{R}) as a function of density for all the state points simulated. The colors correspond to the different isomorphs studied; the same color coding is used in the other figures.}\label{fig6}
\end{figure}

The simulations were performed with the RUMD GPU-based Molecular Dynamics software \cite{rumd}. The $NVT$ ensemble with a Nose-Hoover thermostat was used throughout. All simulations involved 1,000 particles. A shifted-force cutoff was employed with $r_{\rm cut} = 3.5 \sigma$ for the LJ system and $r_{\rm cut} = 2.5 \sigma$ for the two other systems.

\begin{figure}[H]  
  \centering
  \includegraphics[trim = 0mm 0mm 0mm 35mm, clip,width = 0.3 \textwidth]{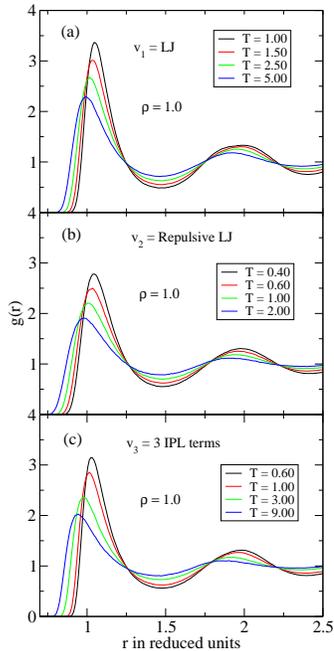}
  \caption{Radial distribution functions of the three systems at unit density and the reference temperatures defining the different isomorphs studied.}\label{fig6a}
\end{figure}

In order to investigate the usefulness also of Eq. (\ref{gest3}), we simulated for each system a number of thermodynamic state points along four isomorphs. The isomorphs were generated using Eq. (\ref{isomch}). According to the isomorph theory, this equation involves a unique function $h(\rho)$ \cite{ing12a}. However, since we are here also interested in investigating possible variations going from one isomorph to another, we determined one function $h(\rho)$ for each isomorph. Thus, following Ref. \cite{boh12} each isomorph was generated via Eq. (\ref{isomch}) from a $h(\rho)$ function calculated at the reference state point defined by $\rho=1$ (in units of $\sigma^{-3}$) and the reference temperature via the correlation functions $\langle\Delta U_n\Delta U\rangle$, in which $\Delta U_n$ is the fluctuation of the $n-$IPL term of $v(r)$ \cite{boh12}.

All LJ state points [pair potential $v_1(r)$] simulated obey $R>0.91$, 
all Repulsive LJ state points [pair potential $v_2(r)$] simulated obey $R>0.99$, and 
all 3-IPL system state points [pair potential $v_3(r)$] simulated obey $R>0.93$ (Fig. 1). Figures \ref{fig6a}(a)-(c) show the radial distribution functions at $\rho=1$ at the reference temperatures from which the isomorphs were generated. Clearly, the state points simulated involve a considerable variation in structure.

\begin{figure}[H]
  \centering
  \includegraphics[trim = 0mm 0mm 0mm 35mm, clip,width = 0.24 \textwidth]{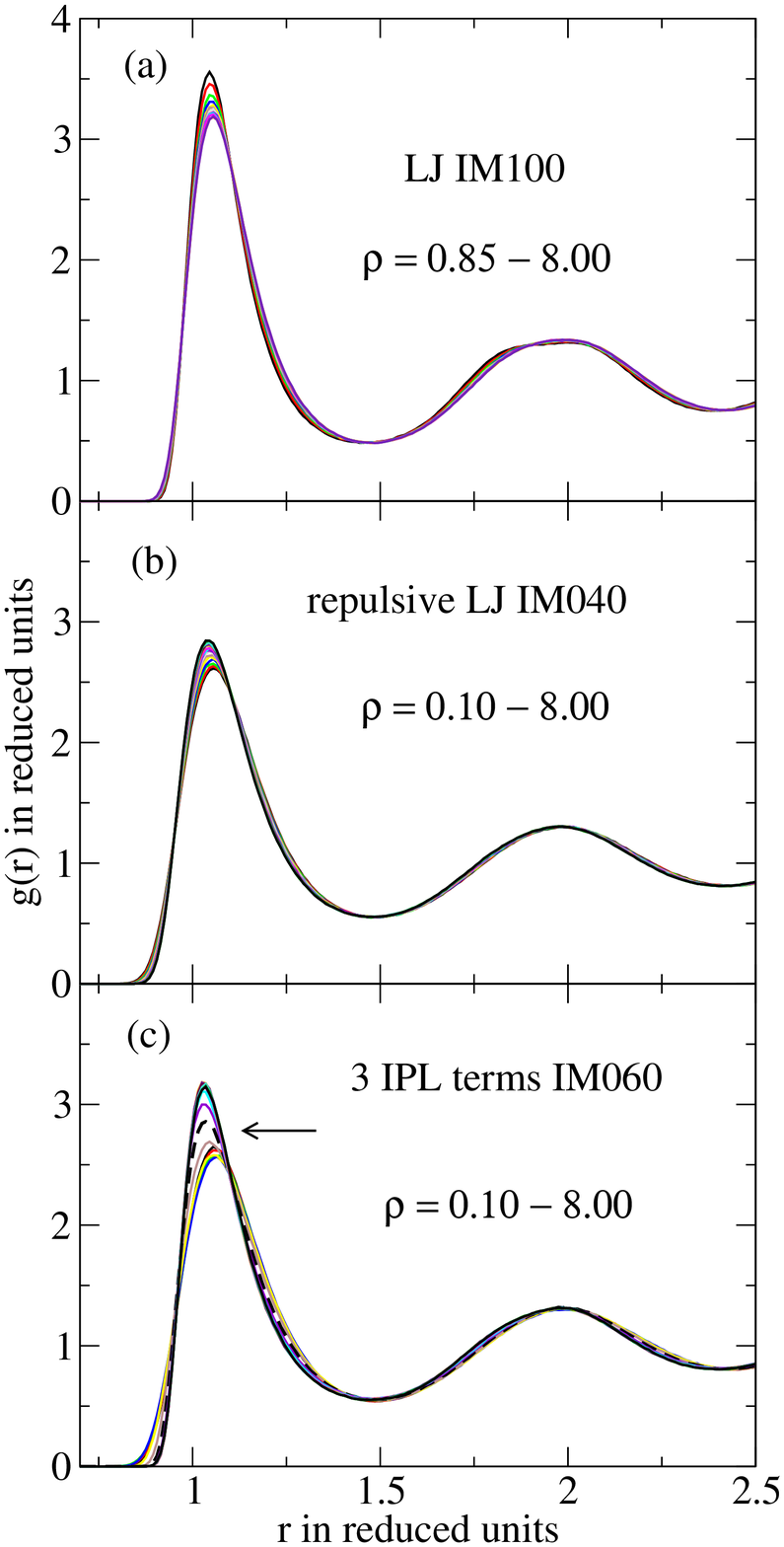}
  \includegraphics[trim = 0mm 0mm 0mm 35mm, clip,width = 0.24 \textwidth]{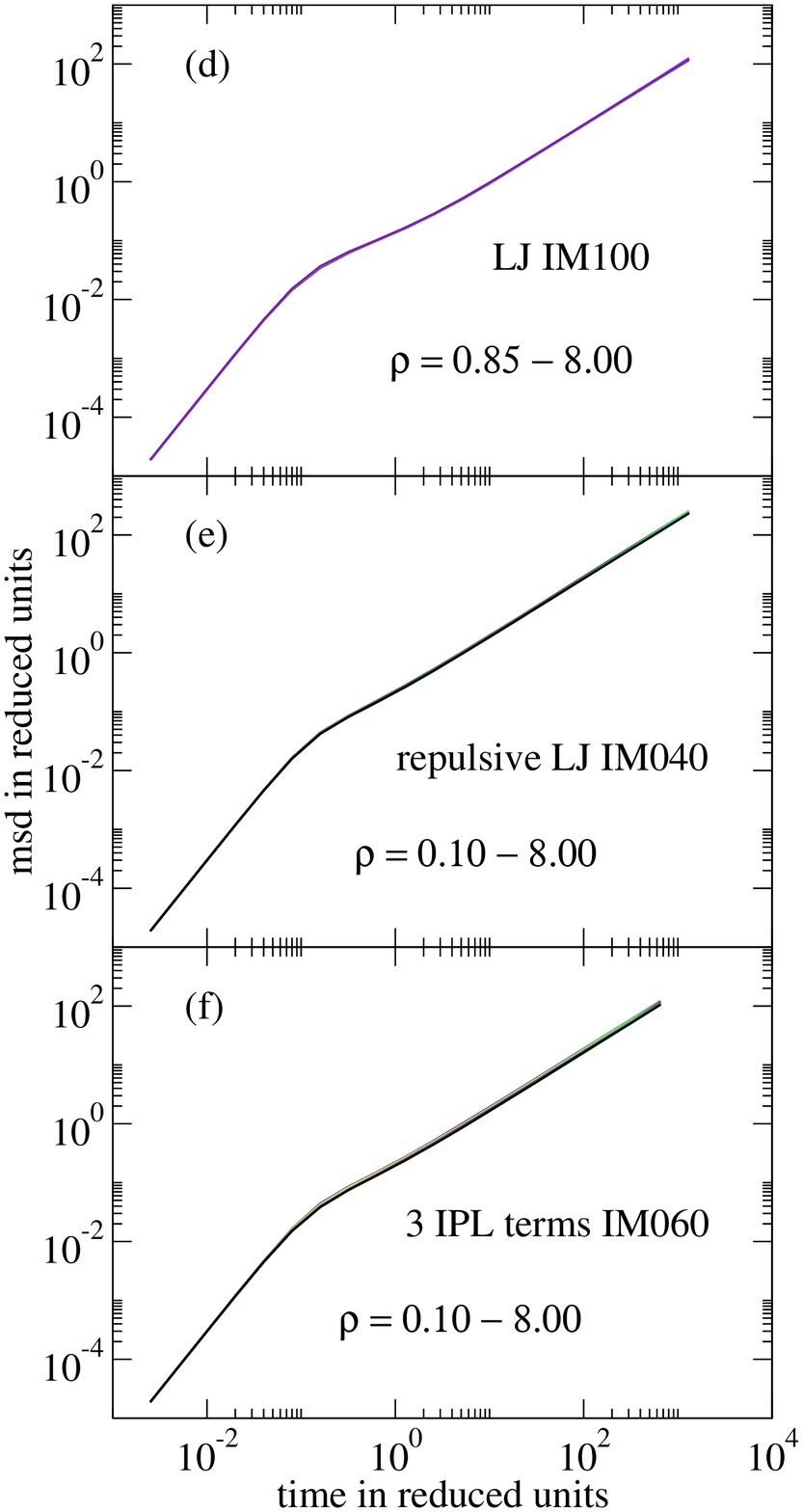}
  \caption{Reduced-unit radial distribution functions $g(r)$ and mean-square displacements along the lowest-temperature isomorphs, one for each of the three systems.
(a) The LJ system has a density range from 0.85 to 8.00, 
(b) the Repulsive LJ system densities ranging from 0.1 up to 8, and 
(c) the 3-IPL system has the same density range. Given the considerable density variation, all three isomorphs are seen to have invariant structure to a quite good approximation. For the 3-IPL system, however, the data bundle into a low- and a high-density set; the broken curve indicated with an arrow is at $\rho = 0.70$ where the system is in between the low- and high-density ``wings'' of the potential. -- The mean-square displacements for the same state points are plotted in reduced units in (d), (e), and (f), respectively -- these are all invariant to a very good approximation.}\label{fig7}
\end{figure}

Before proceeding to test the two proposed expressions for the density-scaling exponent Eqs. (\ref{gest1}) and (\ref{gest3}), we checked the isomorph invariance of structure and dynamics. Figures \ref{fig7}(a)-(c) give the radial distribution functions along the lowest-temperature isomorph studied for each system (the numbers in the legend represent 100 times the reference temperature at density 1.00, so for instance IM040 means that the isomorph was started at  temperature 0.40, in the unit system defined by the $\sigma$s and $\varepsilon$s of Eqs. (\ref{eq:pots1})-(\ref{eq:pots3})). Given the large density range studied, the predicted isomorph invariance of structure and dynamics in reduced units is well obeyed. Notably, the dynamics is more isomorph invariant than the structure, something we have often observed and interpret as follows. The pair potential is not an isomorph invariant, even in reduced units. Thus, since the probability of close encounters is proportional to $\exp(-v(r)/k_BT)$ for $r\rightarrow\infty$, the way $g(r)$ goes to zero for small distances cannot be  isomorph invariant (even in reduced units). This affects the radial distribution function below the first peak of $g(r)$ and often also the peak height, which is what one observes in Fig. 3. Interestingly, we find both here and in previous simulations that this minor deviation from isomorph invariance of structure does not affect the invariance of the dynamics -- collective as well as individual -- an observation that, incidentally, explains the succesful use of the hard-sphere model for reproducing the dynamics of LJ-type liquids.

Another notable point is that for the 3 IPL term system the RDFs cluster into two sets, one for low densities and one for high (Fig. 3(c)). The low-density cluster is where the $n=6$ IPL term dominates, the high-density cluster is where the $n=18$ term dominates. This deviation from perfect isomorph scaling is an expression of violations of quasiuniversality \cite{dyr13}: if the physics of the $n=6$ and the $n=18$ systems were identical, there would be perfect collapse; this is not the case, in part for the above discussed reason that some deviations {\it must} occur.

\section{Comparing the conjectures for the density-scaling exponent to simulation results}\label{compar}

\begin{figure}[H]
  \centering
  \includegraphics[width = 0.3 \textwidth]{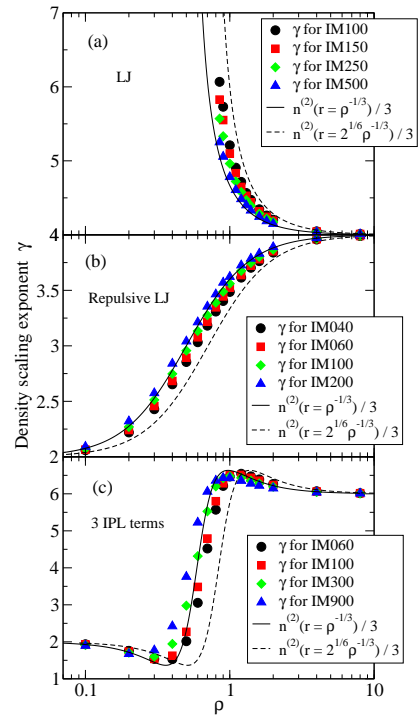}
  \caption{The density-scaling exponent $\gamma$ calculated from Eq. (\ref{gamma_eq}) as a function of the density in simulations (one color for each isomorph) compared to the predictions of Eq. (\ref{gest1}) in which the full curves represent $\Lambda=1$ and the dashed curve $\Lambda=2^{1/6}$ (corresponding to the nearest-neighbor distance of an FCC crystal). }
  \label{fig3}
\end{figure}

We first compare the prediction for the density-scaling exponent Eq. (\ref{gest1}) to simulations for which $\gamma$ was calculated at each state point using Eq. (\ref{gamma_eq}). The results are shown as functions of density in Fig. \ref{fig3} with one color for each isomorph. The full curve gives the prediction of Eq. (\ref{gest1}) with $\Lambda=1$, the dashed curve with $\Lambda=2^{1/6}$. We note the following. Firstly, the three systems have quite different variations of $\gamma$ with density. Secondly, the isomorph-theory prediction that $\gamma$ depends only on density is roughly obeyed, though not 100\%. Thirdly, Eq. (\ref{gest1}) gives a qualitatively correct representation of data for all systems; in particular the significant differences between the three systems are captured by this expression. Overall, the pragmatic choice $\Lambda=1$ works best, but it is should be noted that as temperature is lowered, the data move towards the predicted density-scaling exponent for $\Lambda=2^{1/6}$. This makes good sense, because at lower temperatures the local structure is expected to be more like that of an FCC crystal (for which $\Lambda=2^{1/6}$ \cite{lj37}) than at higher temperatures.

\begin{figure}[H]
  \centering
  \includegraphics[trim = 0mm 0mm 0mm 35mm, clip,width = 0.3 \textwidth]{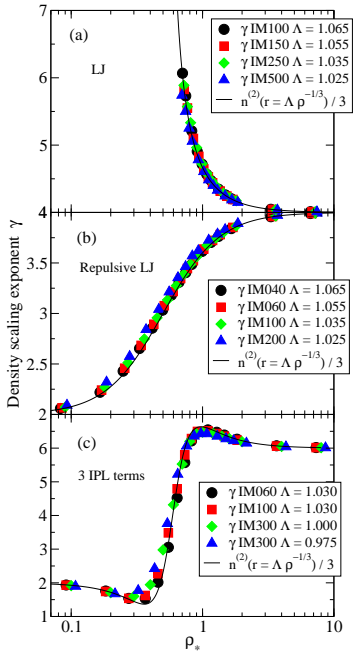}
  \caption{Same simulation data as Fig. \ref{fig3}, but here compared to the prediction of Eq. (\ref{gest3}) (full curve), the expression in which $\Lambda$ for each isomorph is determined from the position of the most likely nearest-neighbor distance. The x-axis variable is defined by $\rho_*\equiv\Lambda^{-3}\rho$. Within our resolution (0.005) the $\Lambda$ values were found to be identical in (a) and (b). }
  \label{fig5}
\end{figure}

We proceed to compare the simulation data to the prediction of the more general Eq. (\ref{gest3}), for which $\Lambda\equiv\Lambda(\Sex)$ determines the most likely nearest-neighbor distance as equal to $\Lambda\rho^{-1/3}$ at the reference temperature of the given isomorph; perfect isomorph invariance of the structure would imply that $\Lambda$ is constant along an isomorph, which indeed applies to a quite good approximation (data not shown). The comparison to Eq. (\ref{gest3}) is shown in Fig. \ref{fig5}. In order to represent Eq. (\ref{gest3}) as a single curve for the different isomorphs the x-axis has been redefined to the quantity $\rho_*\equiv\Lambda^{-3}\rho$. Comparing to Fig. 4 it is clear that Eq. (\ref{gest3}) provides a better fit to data than Eq. (\ref{gest1}).

\section{Discussion}\label{disc}

Our simulations show that Eq. (\ref{gest1}) provides a qualitatively correct analytical estimate of the density-scaling exponent from the pair potential. The simulations  moreover show that the (minor) deviations from the isomorph-theory prediction that $\gamma$ depends only on density to some degree can be rationalized by assuming that the scaling factor converting density to a distance varies slightly from one isomorph to another (Eq. (\ref{gest3})). That such a variation must be allowed for in order to get a more accurate prediction for $\gamma$ is not surprising, given the fact that the isomorph theory is only approximate and that the structure varies  between different isomorphs.

In the rigorous isomorph theory $\gamma$ is given by Eq. (\ref{gheq}) in which $h$ depends only on density. Having in mind the more general expression Eq. (\ref{gest3}), we can generalize Eq. (\ref{gheq}) by proceeding as follows. First one notes that Eq. (\ref{np}) for $p=2$ can be written as $n^{(2)}(r)=-d\ln[r^2v''(r)]/d\ln r$. Thus, since $d\ln r=-d\ln \rho/3$, Eqs. (\ref{gammadef}) and (\ref{gest3}) imply

\be\label{longeq}
\left(\frac{\partial \ln T}{\partial \ln \rho} \right)_{\Sex}
\,=\,\left(\frac{\partial\,\ln[r^2v''(r)]|_{r=\Lambda(\sex)\rho^{-1/3}}}{\partial\ln\rho}\right)_{\Sex}\,.
\ee
Integrating this leads to $\ln T = \left.\ln[r^2v''(r)]\right|_{r=\Lambda(\sex)\rho^{-1/3}} + K(\sex)$ for some function $K(\sex)$. This means that one can write $k_BT=f(\sex)h(\rho,\sex)$ in which $f(\sex)=k_B\exp[K(\sex)]$ and

\be\label{hex}
h(\rho,  \sex) \equiv A  \rho^{-2/3} \left. v''(r)\right|_{r=\Lambda(\sex)\rho^{-1/3}}.
\ee
Here $A$ is an (arbitrary) multiplicative constant; it can be chosen such that $h$ is unity at a particular reference density on a given isomorph. Note that given $\Lambda(\sex)$, which can be determined from a single simulation at the reference temperature of a given isomorph, Eq. (\ref{hex}) provides a convenient way of tracing out the entire isomorph via 

\be
\frac{h(\rho,\sex)}{T}\,=\,{\rm Const.}
\ee
This generalized the isomorph theory's recipe Eq. (\ref{isomch}).

Choosing for now $A=1$, in terms of reduced coordinates \cite{IV} one has 

\be\label{h_rho_S}
\frac{h(\rho,  \sex)}{k_BT}
\,=\,\left.\tilde{v}''(\tilde r)\right|_{\tilde r=\Lambda(\sex)}\,.
\ee
Physically this corresponds to the square of an effective, reduced ``Einstein'' frequency of a single particle pair. It makes good sense that the relevant reduced distance $\Lambda$ at which to evaluate this quantity corresponds to the most likely (reduced) nearest-neighbor distance. We emphasize that isomorph invariance of $\left.\tilde{v}''(\tilde r)\right|_{\tilde r=\Lambda(\sex)}$ is not a trivial consequence of the isomorph theory. In fact, the statement that this expression is isomorph invariant expresses in a concise way the main findings of this paper.

All together we conclude that the scaling properties of monatomic Roskilde-simple liquids are now fairly well understood, although it would be nice to have a simple analytical theory that allows for calculation of the density-scaling exponent with the accuracy of Eq. (\ref{gest3}) without any simulation input. A similarly good understanding applies for neither multicomponent atomic nor molecular Roskilde-simple systems. These are large and important classes of systems that represent important challenges for future work.

\acknowledgments
The centre for viscous liquid dynamics ``Glass and Time'' is sponsored by the Danish National Research Foundation via grant DNRF61.

\end{document}